\documentclass[aps,twocolumn,showpacs]{revtex4}

\usepackage{graphicx}

\def\dIdV{\mathrm{d}I/\mathrm{d}V}
\def\E#1{E_\mathrm{#1}}

\begin{document}

\title{Break-down of the density-of-states description of scanning
 tunneling spectroscopy\\ in supported metal clusters}

\author{Mario De\,Menech$^{1,2}$}
\author{Ulf Saalmann$^1$}
\author{Martin E.\ Garcia$^2$}
\affiliation{$^1$Max--Planck--Institut f\"ur Physik komplexer Systeme\\
  N\"othnitzer Str.\ 38, 01187 Dresden, Germany}
\affiliation{$^2$Theoretische Physik, FB 18, and Center for
 Interdisciplinary Nanostructure Science and\\ Technology,
  Universit{\"a}t Kassel, Heinrich-Plett-Str.\ 40, 34132 Kassel,
  Germany}
\date{\today}

\begin{abstract}\noindent
  Low-temperature scanning tunneling spectroscopy allows to probe the
  electronic properties of clusters at surfaces with unprecedented
  accuracy.  By means of quantum transport theory, using realistic
  tunneling tips, we obtain conductance curves which
  considerably deviate from the cluster's density of states.  Our
  study explains the remarkably small number of peaks in the
  conductance spectra observed in recent experiments.  We demonstrate
  that the unambiguous characterization of the states on the supported
  clusters can be achieved with energy-resolved images, 
   obtained from a theoretical
    analysis which mimics the experimental imaging procedure.
\end{abstract}


\maketitle

The investigation of the properties of atomic clusters is of great
fundamental interest since this class of systems allows to bridge the
gap between single atoms on one side and bulk material on the other
side \cite{cluster}.  
This applies to both free clusters in the gas phase and
clusters deposited at surfaces.  Compared to the gas phase, where
characteristic effects like magic numbers and size dependent
bond-character changes are well studied, the understanding of
supported clusters is much less developed due to the the numerous
technical difficulties the problem poses to both experiment and
theory.  However, there is a growing interest in supported clusters,
mainly for two reasons: firstly, by means of a scanning tunneling
microscope (STM) one is now able to perform single-cluster studies
\cite{nier+99}, much more difficult to achieve in the gas phase
because of the low beam intensity of mass-selected clusters.
Additionally, scanning tunneling spectroscopy (STS) provides valuable
information on the electronic properties, also allowing spatial
resolution, in contrast to the non-local character of photoelectron
spectroscopy \cite{hamo+04}.  Secondly, supported clusters promise to
be efficient catalysts as oxidation experiments with small gold
clusters indicate \cite{vala+98,hesc00}.  STM/STS techniques represent
unique tools to analyze supported clusters and thus to shed light on
the oxidation mechanism.

STS measurements of metallic clusters (platinum \cite{beko+98},
silver \cite{hogr+00}, and gold \cite{baho03,hoba06}) on graphite
surfaces were performed in recent years in order to verify the effect
of the substrate on the electronic structure of the clusters.  The STS
spectra showed peaks which were unambiguously shown to be produced by
cluster states. 
Also, experiments ruled out that for metal clusters at graphite
the observed peaks are due to Coulomb blockade effects
\cite{hogr+00,hoba06}. 
Surprisingly, the number of peaks obtained for the considered
bias-voltage range was considerably smaller than the number of
electronic states expected for the cluster within the
corresponding energy interval \cite{beko+98,hogr+00,baho03,hoba06}.  
These facts clearly indicate
that there are states of the clusters which do not contribute to the
conductance, and that the simple interpretation of the STS spectra
only in terms of the density of states (DOS) \cite{teha8385} is not sufficient.
Moreover, no systematic trend of the peak distribution as a
function of the cluster size could be singled out.  
Besides, repeated measurements of the same cluster
showed variations in the spectra \cite{hoba06}. 
 The blurry picture emerging from these observations calls for a
 theoretical analysis in order to
evaluate the dependence of the conductance spectra on the
details of the STS measurement.

In this Letter we demonstrate that the description of STS spectra only
in terms of the DOS breaks down in the case of metallic clusters on
graphite surfaces.  Furthermore, we show that the interpretation and
understanding of STS spectra requires the careful calculation of the
electron transport through the cluster to the infinite supporting
surface, taking into account a realistic shape for the STM tip.  Our
approach, based on non-equilibrium Green's functions (NEGF), treats
cluster and surface on an equal microscopic footing.  It
allows us to determine the differences between the DOS and
the conductance for different cluster sizes, tip
shapes and surface-tip distances.  We will show that there is a
crucial dependence of the STS spectra on the lateral position of the
tip which should always be kept in mind when discussing experimental
spectra. 
As we report in our simulations, a reliable characterization of
the cluster states should be done by means of energy resolve
imaging, therefore combining spectral information with spatial
resolution.

We performed quantum transport calculations for silver clusters having
up to 233 atoms on a perfect graphite surface.  
We consider the STM scenario in which the tip of the STM is at a distance
of a few {\AA} over the cluster and calculate the tunneling
current $I$ between tip electrode and cluster; all the
components are treated at the atomistic level.  
For the cluster we assume the lattice constant of the Ag fcc
bulk crystal structure and use a Wulff construction
\cite{hawo+99} to define the facets \cite{mesa+06}.   
By means of an effective surface energy for silver and graphite
we can vary the width-to-height ratio of the clusters to match
the experimental value \cite{hogr+00}.  
The lateral position of the cluster on the surface was found to
be of minor importance; the cluster-surface distance was set to
2.5\,\AA, consistent with ab initio calculations for Ag monomers
and dimers on graphite \cite{wabe+03}.   
The platinum electrode above the cluster is either
just a flat ideal surface or a sharp pyramidal
Pt tip mounted on the electrode surface.  The flat electrode is
used for reference calculations; the sharp tip, having a
pyramidal structure of three layers with 1-3-6 atoms,
can be considered as a model of a real STM tip \cite{stm-book}.

The electronic structure and the transport properties of the whole
system, i.\,e.\ graphite surface, silver cluster and platinum
electrode/tip, are calculated by means of an NEGF approach
\cite{da95,brmo+02}.  
The DOS and the current $I$ at a given bias $V$ are calculated
from the Green's functions of the cluster; 
they contain the interaction with the surface and the
electrode/tip via tunneling self-energies \cite{da95}.  
We employ a self-consistent tight-binding model which is
parameterized from density-functional calculations; atomic charge
fluctuations (transfer and polarization) are taken into account
\cite{elpo+98}.  The procedure for calculating equilibrium properties,
like the charge transfer to or from the cluster, and transport
properties, like the current through the cluster, is described
in Ref.~\cite{mesa+06a}.

\begin{figure}[b]
\centering
\includegraphics[width=0.4\textwidth]{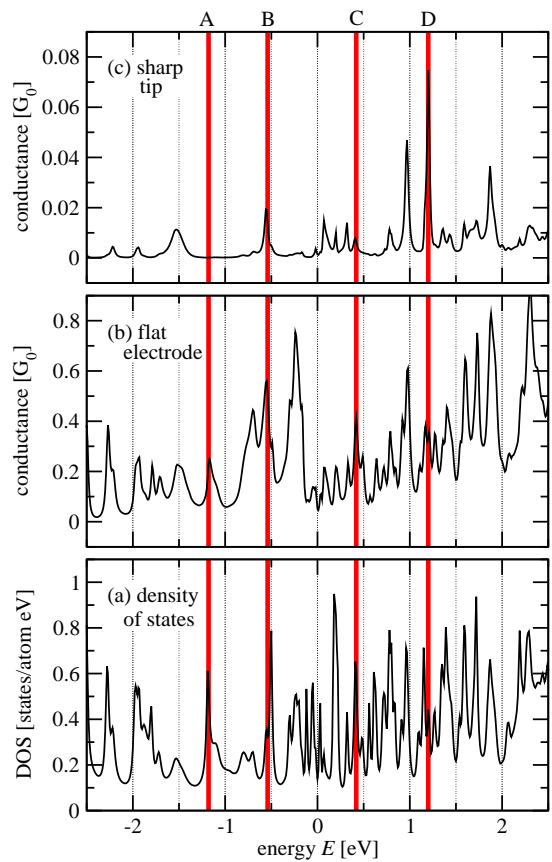}
\caption{(Color online) 
  Density of states (panel a) of an Ag$_{112}$ cluster at
  a graphite surface compared to the conductance spectra
  obtained with flat (b) and sharp (c) platinum electrodes
  as a function of energy, being $E=0$ the Fermi energy.
  G$_{0}=2 e^2/h$ is the quantum unit of conductance.}
\label{fig:compare_flat_sharp_Ag112}
\end{figure}
Figure~\ref{fig:compare_flat_sharp_Ag112} presents a comparison
of the DOS with the conductance $\dIdV$ for  Ag$_{112}$, 
calculated for both tip shapes described above.  
Two different regions can be identified in the DOS of the
supported cluster, based on the mean  level spacing: 
a rich structure with many peaks and level distances
of $\Delta E\approx0.1$\,eV is visible above the Fermi energy
($E\gtrsim0$), while for the occupied states ($E\lesssim0$) we
have a larger spacing, $\Delta E\approx0.5$\,eV.  
The latter value is similar to the energy level spacing measured
with photoelectron spectroscopy for Ag clusters of comparable size,
cf.\ Fig.\,2 of Ref.\,\cite{hamo+04}.  
One should mention that, in the case of supported silver
clusters it is hardly possible to assign certain levels to the
peaks as it is done for free clusters \cite{hamo+04} or
supported fullerenes \cite{mesa+06a}.  
By calculating the DOS of the clusters placed at different
heights over the surface, we observed that, as the clusters
approach the surface, the levels are not only broadened but are
also shifted to higher or lower energies in a complicated way.

\begin{figure*}[hbt]
\centering
\includegraphics[width=0.9\textwidth]{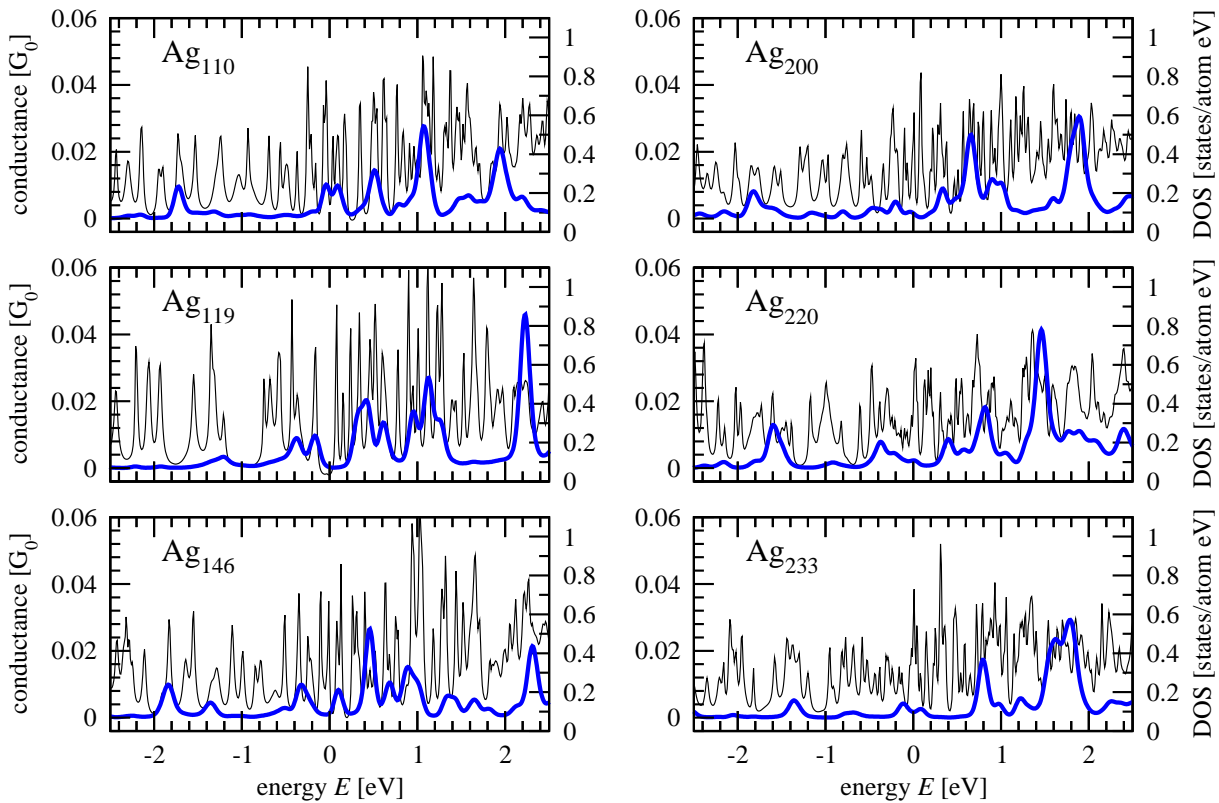}
\caption{(Color online) 
Conductance (thick blue lines, left axis) through silver
clusters of various sizes on a graphite surface; the sharp Pt
conductance curves are convoluted with a Gaussian having 100\,meV full
width at half maximum.  For comparison the density of states (thin
lines, right axis) is shown as well. 
G$_{0}=2 e^2/h$ is the quantum unit of conductance.}
\label{fig:clusatsu}
\end{figure*}
Next, we compare the DOS to the conductance patterns for the flat
and the sharp tips, taken as representatives of two limiting
cases. 
Panel (b) of Fig.\,\ref{fig:compare_flat_sharp_Ag112} shows
$\dIdV$ for the flat electrode.  
The relative peak height changes compared to the DOS; some
peaks are reduced, for example the peak A at $\E{A}{=}-1.18$\,eV,
while others are enhanced, for example at $\E{B}{=}-0.54$\,eV.
Despite these variations, the overall picture is largely preserved and
we conclude that the conductance measured with the artificially flat
electrode would give a reasonable picture of the DOS of the cluster.
The situation changes drastically for the sharp tip, as shown in panel
(c) of Fig.\,\ref{fig:compare_flat_sharp_Ag112}.
Due to the reduced number of atoms of the probe participating to
the tunneling process, the conductance drops by more than an order of
magnitude (in fact we observed that for the flat electrode the
average conductance scales directly with the nuber of atoms in the top
facet of the cluster).
More remarkably, a few peaks completely disappear, for example
the one labeled A, while many others are strongly suppressed.  
The discrepancy between the spectra obtained with the two
kinds of electrodes can be understood as follows.  While the flat
electrode signal receives contribution from all atoms on the top
facet, representing the average DOS, the sharp tip probes the local
DOS in a small region, and the resulting conductance will be affected
by the spatial dependence of the specific state (see below).
Independently of the detailed structure of the probe in the
experiments, we expect that the measured conductances should be
related to our sharp-tip results.  Any atom sticking out of the tip
will carry almost all current in the tunneling regime.

We have extended the study to other cluster sizes and present
six of them in Fig.\,\ref{fig:clusatsu}.
In each of the graphs we compare the
sharp-tip conductance with the DOS of the supported cluster.  As in
Fig.\,\ref{fig:compare_flat_sharp_Ag112} we placed the tip centrally
over the cluster.  Additionally, the conductance curves were
convoluted with a Gaussian function of 100\,meV full width at half
maximum, in order to reproduce the accuracy of experimental
low-temperature measurements \cite{hoba06}.  Even if the detailed
shape of the curves differs from one cluster size to the next one,
the general message is the same: the conductance is much less
structured than the corresponding DOS.  Thus we conclude that the
experimentally found larger level spacing in supported cluster
\cite{beko+98,hogr+00,baho03,hoba06}, if compared to the gas phase
\cite{hamo+04}, is due to the local and selective character of the
transport measurement. 
However, in accordance with the experimental
observation \cite{hoba06}, positions and widths of the conductance
peaks in Fig.\,\ref{fig:clusatsu} do not show any trend with the
cluster size.

So far we have described situations in which the sharp tip was placed
over the cluster center.  Moving the tip laterally from the central
position may change the observed conductance spectra substantially.
Energy resolved maps are obtained by projecting the local
conductance on a constant current isosurface at particular
energies $E_X$. 
The constraint of a constant current isosurface is necessary to
compare conductance values measured at different positions over
the sample. 
This point represents the main difficulty in the calculations,
since tip trajectory invariably describes a complicated surface.

\begin{figure*}[t]
\centering
\includegraphics[width=0.45\textwidth,angle=270]{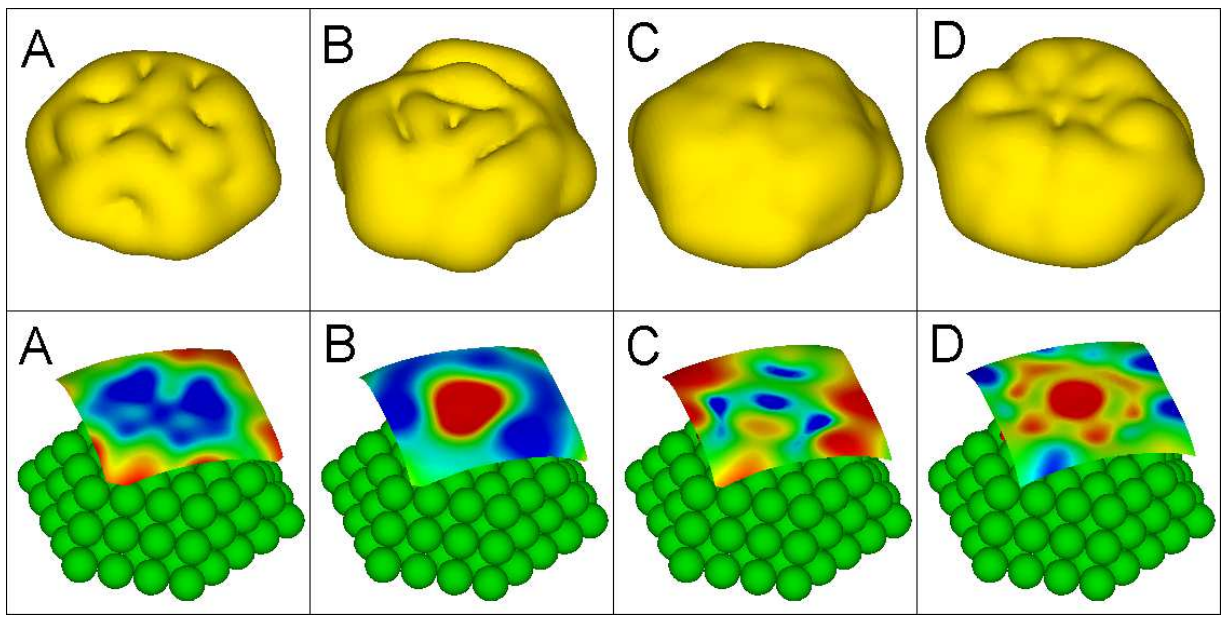}
\caption{(Color online) Constant-density isosurfaces (upper row) 
  and conductance maps (lower row) 
  for four selected states of Ag$_{112}$.
  The four peaks are marked by red lines in
  Fig.\,\ref{fig:compare_flat_sharp_Ag112}; 
  their energies are 
  $\E{A}{=}{-}1.18$\,eV, $\E{B}{=}{-}0.54$\,eV,
  $\E{C}{=}{+}0.42$\,eV and $\E{D}{=}{+}1.20$\,eV, respectively.
  The conductance maps are presented as contour plots on
  constant-current isosurfaces (see text);
  red (light grey) = high conductance, blue (dark grey) = low conductance.
  The side length of the squares is 14.5\,\AA.}
\label{fig:maps}
\end{figure*}

In Fig.\,~\ref{fig:maps} we report such maps for the states
labeled A--D in Fig.\,\ref{fig:compare_flat_sharp_Ag112}.  
Figure~\ref{fig:maps} shows for each of the chosen energies
the constant-density isosurface (upper row) and the conductance map
(lower row).
The former has been obtained by integrating the lesser Green's
function over a small energy interval of 20\,meV centered
around the energies $E_\mathrm{A}$, $E_\mathrm{B}$,
$E_\mathrm{C}$, and $E_\mathrm{D}$ in order to get the charge density
in terms of the local atomic orbitals.  The constant-density
isosurfaces show that all four states are delocalized over the whole
clusters.  However, their shapes differ, in particular at the upper
facet.

A much clearer picture of these different structures emerges from the
conductance maps, shown in the lower row of
Fig.\,\ref{fig:maps}.  
These maps display as a contour plot the conductance
$\dIdV(x,y,z,E_X)|_{I=I_0}$ on a constant-current isosurface $z=z(x,y)$   
 determined
by the implicit equation $I(z;x,y)\stackrel{!}{=}I_0$.
In the calculations, the tip was moved on a three-dimensional
finite-element mesh with a grid spacing of 0.5\,\AA.  
At each grid point the current and the conductance were
computed, and the constant-current isosurface was obtained by
interpolation.   
Having this three-dimensional topographic image of the
cluster we render conductance maps by showing the calculated $\dIdV$
values with a colored contour plot on it.  The obtained patterns are
dominated by the three-fold symmetry of the clusters, which is,
however, disturbed by the coupling to the supporting surface.  The
central minima 
(blue spots in online version, black spots in print version) 
of the patterns A and C clearly reveal the reason for the
disappearance of the corresponding peaks in the measurement with
the tip over the centre, cf.\ Fig.\,\ref{fig:compare_flat_sharp_Ag112}.  
Furthermore, it becomes clear that placing the tip a few {\AA}
off the center leads to considerable changes in the conductance
of these peaks.  
This may explain the observed discrepancies in the spectra
obtained from measurements performed on the same cluster but at
different times \cite{hoba06}.

Summarizing, we have shown that recent data from STS measurement of
metallic clusters at surfaces can be understood only in terms of
non-equilibrium transport calculations.  The calculated current and
conductance, respectively, depend crucially on the
tip shape.  A sharp tip is very selective, which explains the low
number of states seen in the experiment compared to the richly
structured DOS of the cluster.  This selectivity depends strongly on
the tip position, which leads to the requirement of a more complete
experimental characterization based on energy-resolved imaging of
the electronic states on the cluster \cite{lugr+0304}.

We acknowledge financial support by the Deutsche Forschungsgemeinschaft
through the priority program SPP 1153 ``Clusters in contact with surfaces:
Electronic structure and magnetism''.


\begin{thebibliography}{10}
\bibitem{cluster}
  A.~W. Castleman, Jr.\ and P.~Jena, Proc. Natl. Acad. Sci. 
  {\bf 103}, 10552 (2006),
  special issue on {\em Cluster Chemistry and Dynamics}.
\bibitem{nier+99}
  N.~Nilius, N.~Ernst, and H.-J. Freund, Phys. Rev. Lett. {\bf 84}, 3994
  (2000).
\bibitem{hamo+04}
  H.~H{\"a}kkinen, M.~Moseler, O.~Kostko, N.~Morgner, M.~Astruc Hoffmann, and
  B.~von Issendorff, Phys. Rev. Lett. {\bf 93}, 093401 (2004).
\bibitem{vala+98}
  M.~Valden, X.~Lai, and D.~W. Goodman, Science {\bf 281}, 1647 (1998).
\bibitem{hesc00}
  U.~Heiz and W.-D. Schneider, J. Phys. D {\bf 33}, R\,85 (2000).
\bibitem{brfe+96}
  K.~Bromann, C.~F{\'e}lix, H.~Brune, W.~Harbich, R.~Monot, J.~Buttet, and
  K.~Kern, Science {\bf 274}, 956 (1996).
\bibitem{irbo+06}
  T.~Irawan, D.~Boecker, F.~Ghaleh, C.~Yin, B.~von Issendorff, and H.~H{\"o}vel,
  Appl. Phys. A {\bf 82}, 81 (2006).
\bibitem{beko+98}
  A.~Bettac, L.~Koller, V.~Rank, and K.~H. Meiwes-Broer, Surf. Sci. {\bf 404},
  475 (1998).
\bibitem{hogr+00}
  H.~H{\"o}vel, B.~Grimm, M.~B{\"o}decker, K.~Fieger, and B.~Reihl, Surf. Sci.
  {\bf 463}, L\,603 (2000).
\bibitem{baho03}
  I.~Barke and H.~H{\"o}vel, Phys. Rev. Lett. {\bf 90}, 166801 (2003).
\bibitem{hoba06}
  H.~H{\"o}vel and I.~Barke, Prog. Surf. Sci. {\bf 81}, 53 (2006).
\bibitem{teha8385} 
  J.~Tersoff and D.~R. Hamann, Phys. Rev. Lett. {\bf 50}, 1998 (1983); 
  Phys. Rev. B {\bf 31}, 805 (1985).
\bibitem{hawo+99}
  K.~H{\o}rup Hansen, T.~Worren, S.~Stempel, E.~L{\ae}gsgaard, M.~B{\"a}umer,
  H.-J. Freund, F.~Besenbacher, and I.~Stensgaard, Phys. Rev. Lett. {\bf 83},
  4120 (1999).
\bibitem{mesa+06}
  M.~{De\,Menech}, U.~Saalmann, and M.~E. Garcia, Appl. Phys. A {\bf 82}, 113
  (2006).
\bibitem{wabe+03}
  G.~M. Wang, J.~J. BelBruno, S.~D. Kenny and R.~Smith, Surf. Sci.
  {\bf 541}, 91 (2003).
\bibitem{stm-book}
  M.~Tsukadsa et al., chapter 5 of {\em Scanning Tunneling Microscopy III}, 
  R.~Wiesendanger and H.-J.~G{\"u}ntherodt (eds.) Springer Berlin 1996.
\bibitem{da95}
  S.~Datta,
  \newblock  {\em Electronic Tranpsort in Mesoscopic Systems}.
  \newblock Cambridge University Press 1995.
\bibitem{brmo+02}
  M.~Brandbyge, J.-L. Mozos, P.~Ordej\'on, J.~Taylor, and K.~Stokbro, Phys.
  Rev. B {\bf 65}, 165401 (2002).
\bibitem{elpo+98}
  M.~Elstner, D.~Porezag, G.~Jungnickel, J.~Elsner, M.~Haugk, T.~Frauenheim,
  S.~Suhai, and G.~Seifert, Phys. Rev. B {\bf 58}, 7260 (1998).
\bibitem{mesa+06a}
  M.~De\,Menech, U.~Saalmann, and M.~E. Garcia, Phys. Rev. B {\bf 73}, 155407
  (2006).
\bibitem{lugr+0304}
  Energy-resolved imaging has already been demonstrated
  experimentally for C$_{60}$ in:
  X.~Lu, M.~Grobis, K.~H. Khoo, S.~G. Louie,  and M.~F. Crommie,  
  Phys. Rev. Lett. {\bf 90}, 096802 (2003);
  Phys. Rev. B {\bf 70}, 115418 (2004).
\end{thebibliography}
\end{document}